\documentclass[prl, twocolumn, superscriptaddress,10pt]{revtex4-1}
\usepackage[utf8]{inputenc}
\usepackage[T1]{fontenc}
\usepackage{amsmath}
\usepackage[english]{babel}
\usepackage[pdftex]{graphicx}
\usepackage{color}
\usepackage{ulem}
\usepackage{comment}
\usepackage{booktabs}
\usepackage{siunitx}

\usepackage[sort&compress]{natbib}
\usepackage{hyperref}
\DeclareUnicodeCharacter{0301}{\'{}}
\usepackage{amsmath}
\usepackage{amssymb}
\usepackage{mathtools}
\usepackage{braket}
\usepackage{bbold}

\begin{document}

\title{Experimental realization of twofold topological phase transition induced by long-range hoppings in one-dimensional chains}

\author{J. A. Franco-Medrano}
\affiliation{Departamento de Ciencias B\'asicas, Universidad Aut\'onoma Metropolitana-Azcapotzalco, Av. San Pablo 420, Col. Nueva el Rosario, 02128, Azcapotzalco, Ciudad de M\'exico, M\'exico.}

\author{B. Manjarrez-Monta\~nez}
\affiliation{Instituto de Ciencias F\'isicas, Universidad Nacional Aut\'onoma de M\'exico, Cuernavaca, Mor., M\'exico}

\author{E. Flores-Olmedo}
\affiliation{Departamento de Ciencias B\'asicas, Universidad Aut\'onoma Metropolitana-Azcapotzalco, Av. San Pablo 420, Col. Nueva el Rosario, 02128, Azcapotzalco, Ciudad de M\'exico, M\'exico.}
\author{G. B\'aez}
\affiliation{Departamento de Ciencias B\'asicas, Universidad Aut\'onoma Metropolitana-Azcapotzalco, Av. San Pablo 420, Col. Nueva el Rosario, 02128, Azcapotzalco, Ciudad de M\'exico, M\'exico.}

\author{A. M. Mart\'inez-Arg\"uello}
\affiliation{Instituto de F\'isica, Benemérita Universidad Autónoma de Puebla, C. P. 72570 Puebla, Puebla, M\'exico}

\author{Y. Betancur-Ocampo}
\affiliation{Instituto de F\'isica, Universidad Nacional Aut\'onoma de M\'exico, Ciudad de México, M\'exico}

\author{R. A. M\'endez-S\'anchez}
\email{mendez@icf.unam.mx}
\affiliation{Instituto de Ciencias F\'isicas, Universidad Nacional Aut\'onoma de M\'exico, Cuernavaca, Mor., M\'exico}

\begin{abstract}
Long-range hoppings offer a powerful mechanism to engineer exotic topological phases beyond conventional tight-binding models. Here, we experimentally demonstrate a twofold topological phase transition driven by third-nearest-neighbor couplings in a one-dimensional phononic chain that emulates cis-polyacetylene. Tunable hopping parameters in highly structured aluminum plates allow us to observe the emergence of Dirac points, a valley pseudospin structure, and topological edge states. The experimental results are supported by non-equilibrium Green's function transport calculations. Our findings confirm the predicted long-range phase diagram and establish phononic metamaterials as a versatile platform for exploring topological phenomena beyond the Su-Schrieffer-Heeger paradigm.

\end{abstract}

\maketitle
Recently, metamaterials have emerged as a versatile platform for exploring topological physics  \cite{Freeney2022,Belopolski2017}.  This field has received increasing attention since there are many open questions and experimental challenges. The key property of these materials is the existence of edge states in an otherwise insulator system \cite{Torrent2013,Suesstrunk2015,Pal2016,Obana2019}. These edge states are immune to changes in a wide range of parameters since they are protected by topology~\cite{Asboth2016}. When this protection is {lifted}, a topological phase transition occurs: in most cases, a Dirac cone emerges at the critical value of the transition, and for larger parameters, a trivial {insulator} behavior is observed \cite{Torrent2013,Asboth2016,Pal2016,Rufo2019}.

The Su–Schrieffer–Heeger (SSH) model, developed to study polymeric chains, captures the simplest topological phase transition driven by alternating nearest-neighbor hoppings between two equal sites in a one-dimensional tight-binding chain \cite{Su1979,Heeger1981,Fradkin1983,Heeger1988,Li2014,Asboth2016}. 
Up to now, generalizations of this model include non-Hermitian topology \cite{Lieu2018,Okuma2023,Manda2024}, topological edge states \cite{Xie2019,Kim2020,Ahmadi2020,Xu2020,Li2022,CaceresAravena2022,Du2025,DalPoggetto2024,Tang2025,Guo2025,Du2025,Bomantara2026}, domain walls \cite{Han2020,Sharma2024}, topological pumping \cite{Wang2026}, long-range hoppings \cite{Zhang2022,BetancurOcampo2024,AbarcaRamirez2026,Khan2026,Sharma2026}, nonlinear effects \cite{Manda2023,Anglart2025}, gate-tunable phase transition \cite{Splitthoff2024,Jakubsky2025,Jakubsky2026}, and Klein tunneling \cite{BetancurOcampo2024a}. 
Despite detailed theoretical and numerical analyses, experimental realizations of long-range hopping topological transitions remain scarce. 
Artificial platforms, ranging from photonic waveguides to mechanical lattices, enable precise control of coupling strengths but typically focus on nearest-neighbor interactions~\cite{CaceresAravena2022,FrancoVillafane2013,Stegmann2017,RamirezRamirez2020,MartinezArgueello2022}. 
Recently, we found a twofold topological transition driven by third-nearest-neighbor hoppings \cite{BetancurOcampo2024}. 
In contrast to the SSH transition, which goes from a trivial to a topological phase, the new twofold topological transition goes from a non-trivial phase to another one with an opposite winding number.

In this letter, we present experimental evidence of a twofold topological phase transition using cis-polyacetylene coupled-resonator phononic molecules (CRPnMs) operating in the ultrasound regime. 
The systems were engineered by coupling resonators through finite phononic crystals (FPnCs). 
The emergent vibrations of such systems obey a tight-binding model similar to that of condensed matter physics. 
By tailoring the geometrical properties of the FPnCs we control the ratio between first- and third-nearest-neighbor couplings, the parameter governing the transition. 
Finite versions of those structures exhibit the emergence of edge states associated with the nontrivial topological invariant. 
Three CRPnMs were fabricated with coupling parameters corresponding to the Dirac point and to each adjacent topological phase. Resonant ultrasound spectroscopy measurements reveal the Dirac cone, the valley pseudospin structure, and the associated topologically protected modes whose wave amplitudes decay exponentially toward the center of the chain. 
This confirms the existence of the twofold topological phase transition. 
The experimental results are also corroborated by Non-Equilibrium Green's Function (NEGF) transport calculations.

\begin{figure}[tbp]
    \centering
    \includegraphics[width=\linewidth]{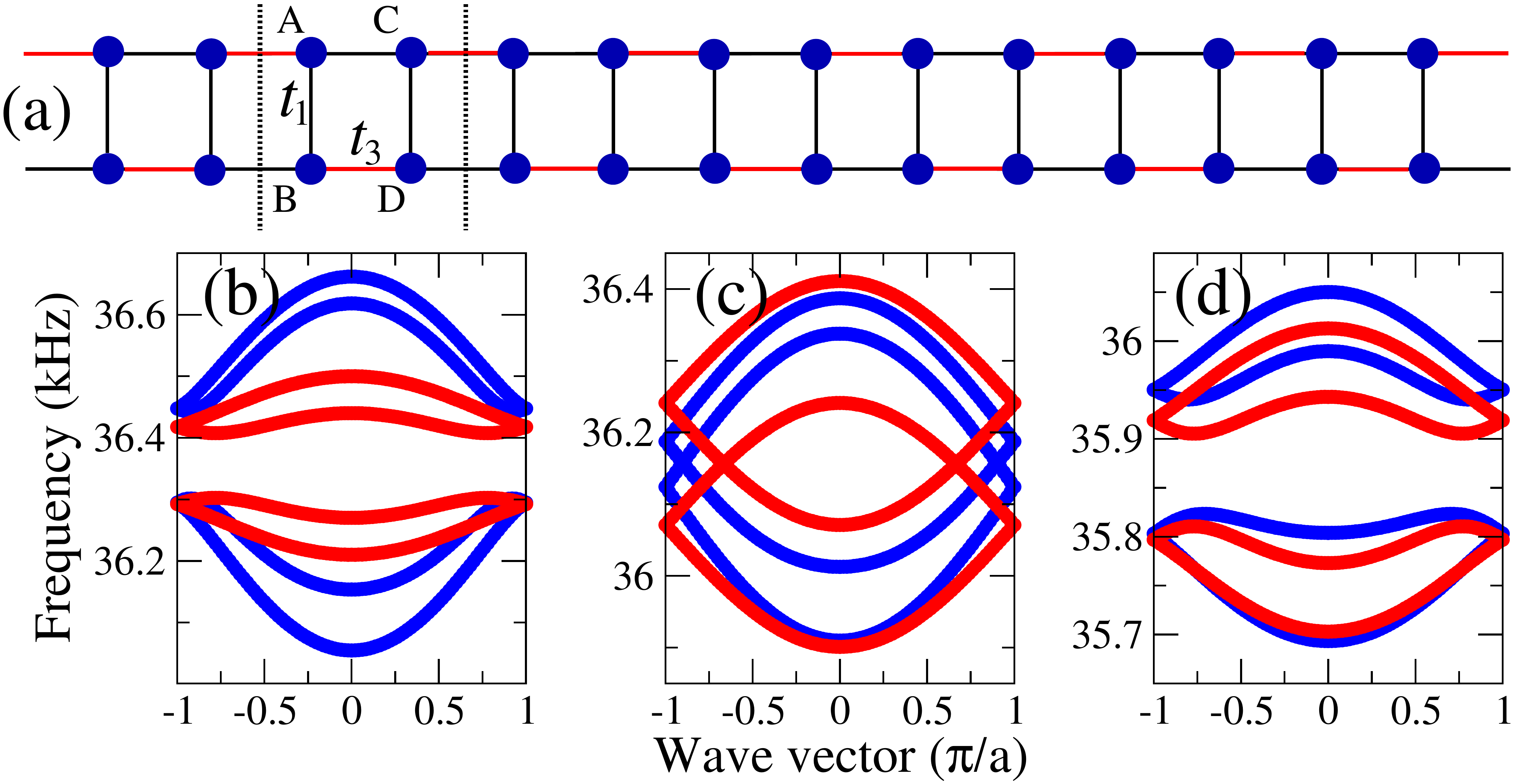}
    \caption{(a) Finite generalized SSH chain with 28 sites. 
    The blue circles depict the sites that are connected through first- (black line) and third- (red line) nearest-neighbor hoppings; sites A, B, C, and D correspond to the non-equivalent positions within the unit cell (vertical dotted lines). The band structure of the unit cell: the red lines represent the theoretical model, and the blue line corresponds to the numerical simulations using COMSOL Multiphysics. (b) $\rho<1$, (c)~$\rho =1$ and (d) $\rho >1$. 
    }
    \label{BandStructure}
\end{figure}

The system, depicted in Fig.~\ref{BandStructure}(a), consists of sites in a one-dimensional chain connected through first- ($t_1$) and third- ($t_3$) nearest-neighbor hoppings indicated by black and red lines, respectively. 
The tight-binding Bloch Hamiltonian for this chain is 

\begin{equation}
H(k)= \begin{pmatrix}
\epsilon_0 & t_1 & 0 & g_1^*(k)\\
t_1 & \epsilon_0 & g_2^*(k) & 0 \\
0 &  g_2(k) &\epsilon_0  & t_1 \\
g_1(k) & 0 & t_1 & \epsilon_0 
\end{pmatrix},
\label{Eq.Hamiltoninan}
\end{equation}
where $\epsilon_0$ is the onsite energy, $g_1(k)=t_3+t_1\textrm{e}^{ika}$ and $g_2(k)=t_1+t_3\textrm{e}^{ika}$, with $a$ the lattice constant and $k$ the wave vector. 

The band structure associated to the previous Hamiltonian is shown in Fig.~\ref{BandStructure} for different values of {$\rho=t_1/t_3$}.  As this ratio increases from zero, a gap opens (a first nontrivial phase, winding number $w=-1$) and then closes at $\rho=1$, forming two non-equivalent Dirac cones, which corresponds to pseudospin-valley structure such as that of graphene. Then the gap reopens at $\rho>1$ with opposite winding number $w=+1$, indicating a second nontrivial phase \cite{BetancurOcampo2024}. 

The Hamiltonian in Eq. \eqref{Eq.Hamiltoninan} is related to the one given in Ref~\cite{BetancurOcampo2024} through a similarity transformation. Such a Bloch Hamiltonian can be reduced to a two-level model around the center of the first Brillouin zone and the low energy regime
\begin{equation}
    H_\mathrm{eff}(q)=\left(\Delta +\frac{q^2}{2m}\right)\,\sigma_x+v q\,\sigma_y,
\end{equation}
where $\Delta = t_3$ changes sign at the transition, and $1/2m$ breaks mass symmetry conjugation \cite{BetancurOcampo2024}. The effective parameters are related to the first and third hopping: $1/2m = (2t_3+3t_1/8)a^2$ for the mass and $v = (2t_3-3t_1/2)a$ for the Fermi velocity. 

The topological phases show up also in finite cis-polyacetylene with site energy $\epsilon_{0}$ and first and third nearest-neighbor hoppings, $t_{1}$ and $t_{3}$. The corresponding tight-binding Hamiltonian $H_{\mathrm{fc}}$ for the finite chain has the Toeplitz matrix form~\cite{BetancurOcampo2024}
\begin{align}
\label{eq:HCis}
H_\textrm{fc}(t_{1},t_{3}) = \qquad \qquad \qquad \qquad \qquad \qquad \qquad \qquad \qquad &\nonumber\\
\left(\begin{array}{c c c c c}
h_\textrm{cp}(t_{1},t_{3})  & C(t_{1},t_{3}) & 0 & \cdots & 0  \\
C^\dagger(t_{1},t_{3}) & h_\textrm{cp}(t_{1},t_{3}) & C(t_{1},t_{3})  & 0 & \cdot\\
0 & C^\dagger(t_{1},t_{3}) & \cdot & \cdot & \cdot \\
\vdots & \vdots &  \vdots & \vdots & C(t_{1},t_{3}) \\
0 & 0 & 0 & C^\dagger(t_{1},t_{3}) & h_\textrm{cp}(t_{1},t_{3})
\end{array}\right) , &
\end{align}

\noindent with the matrix blocks

\begin{equation}
h_\textrm{cp}(t_{1},t_{3}) = \left(\begin{array}{cccc}
\epsilon_0 & t_{1} & 0 & t_{3}  \\
t_{1} & \epsilon_0 & t_{1} & 0  \\
0 & t_{1} & \epsilon_0 & t_{1}  \\
t_{3} & 0 & t_{1} & \epsilon_0  
\end{array}\right) ,
\end{equation}
representing the Hamiltonian of each unit cell, and the coupling matrix between cells
\begin{equation}
C(t_{1},t_{3}) = \left(\begin{array}{cccc}
0 & 0 & 0 & 0  \\
0 & 0 & 0 & 0  \\
0 & t_{3} & 0 & 0  \\
t_{1} & 0 & 0 & 0  
\end{array}\right). 
\end{equation}

A well-established approach to transport through molecular systems is provided by NEGF formalism, where the key mathematical entity is the frequency-dependent Green's function~\cite{Datta1995},
\begin{equation}
G(\epsilon) = [\epsilon \cdot \mathbb{1} - H_{\mathrm{fc}} - \Sigma_{i} -\Sigma_{j}]^{-1}
\label{eq:Green}
\end{equation}
which is constructed from the Hamiltonian of the closed system, $H_{\mathrm{fc}}$ in Eq. \eqref{eq:HCis}, with self-energy contacts $\Sigma$ attached at the site $k$, $\Sigma_{k} = -\mathrm{i} \gamma |k\rangle \langle k|$, with broadening or coupling strength to the outside $\gamma$. 

From equation~(\ref{eq:Green}), the probability amplitude between sites $i$ and $j$, the positions where the exciter and the detector on the artificial chain are located, is given by $G_{ij}(f)$, which is related to the magnitude $A(f)$ and phase $\phi(f)$ of a measured signal between those positions as
\begin{equation}
\label{eq:Quantities}
A_{ij}(\epsilon) = B\, \mathrm{Re}\, G_{ij}(\epsilon) + C \,\, \mathrm{and} \,\,\phi_{ij}(\epsilon) = D\, \mathrm{Im}\, G_{ij}(\epsilon) + F.
\end{equation}
For convenience, $A_{ij}(\epsilon)$ and $\phi_{ij}(\epsilon)$ are rescaled by the constants $B, C, D,$ and $F$, since in our experiments the signals are measured in arbitrary units.
As it will be seen below, Eqs.~(\refeq{Eq.Hamiltoninan}) to~(\refeq{eq:Quantities}) are also valid for artificial molecules by replacing energies with frequencies. 

\begin{figure}[h!]
    \centering
    \includegraphics[width=\linewidth]{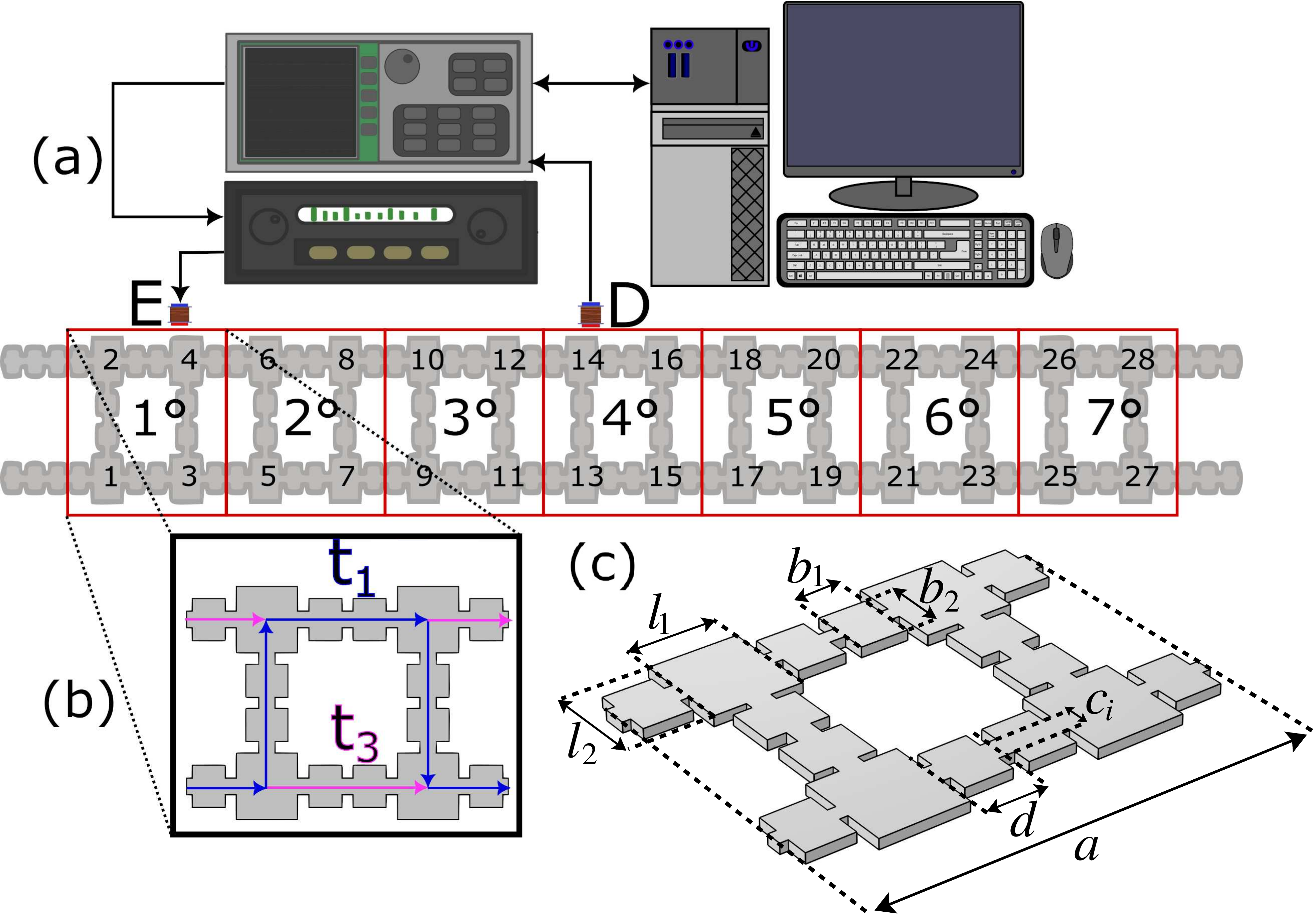}
    \caption{(a) Experimental setup based on contactless electromagnetic-acoustic excitation (E) and detection (D) combined with a vector network analyzer. To optimize selective excitation and detection of the elastic response, the transducers were located near the region of maximum deformation of the in-plane vibrational mode. (b) Elastic analog of the interaction long-range \textit{cis}-polyacetylene chain composed of seven finite unit cells. Adjacent resonators are connected through two distinct phononic crystals emulating the hopping parameters $t_1$ and $t_2$. (c) Unit cell of the phononic molecule with $l_1=55$ mm, $l_2=60$ mm, $a=27$ cm, $b_1=30$ mm, $b_2=38$ mm, $d=40$ mm, $c_1=15$ mm and $c_3=13, 15, 16$ mm.}
    \label{setup}
\end{figure}

Aluminum resonators coupled through phononic-crystal waveguides, that allow transmission only through an emergent miniband, were proposed as an elastic-wave implementation to study topological phase transitions~\cite{BetancurOcampo2024}. 
Such systems, known as coupled-resonator phononic molecules, obey the tight-binding model with hopping frequencies proportional to evanescent overlap, offering independent control of $t_1$ and $t_3$ by varying the geometrical properties of the FPnCs. 

The phononic cis-polyacetylene with long-range interactions consists of seven unit cells shown in Fig.~\ref{setup}(a); each cell contains four sites, as seen in Fig.\ref{setup}(b). 
The sites, square resonators, were designed to exhibit an in-plane resonant mode within the frequency range of 35–37 kHz.

\begin{figure*}[tbp]
    \centering
    \includegraphics[width=0.99\linewidth]{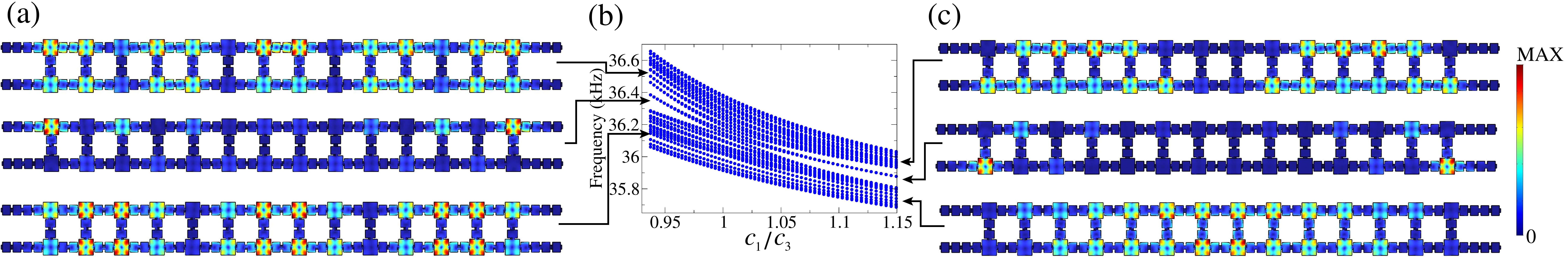}
    \caption{(a) Numerical results for system I, first topological phase. Two bulk states are depicted as well as a topological state located within the gap, which exhibits exponential decay towards the center of the chain.
    (b) Frequency spectrum and normal wave amplitudes of a finite coupled-resonator phononic cis-polyacetylene with first- and third- nearest-neighbor hoppings, as a function of the ratio $c_1/c_3$. (c) Numerical results for system III, second topological phase. Two bulk states are shown along with a topological state located within the gap.}
    \label{fig:placeholder}
\end{figure*}

The unit cell of the chain is depicted in Fig.~\ref{setup}(c). The FPnCs consist of two plates and three cuboids where the width $c_i$ of the cuboids ($i=1,3$) determines the strength of the first- ($t_1$) and third- ($t_3$) nearest-neighbor hoppings, respectively. The tight-binding regime for elastic waves emerges when a resonance of a square plate (site) falls within the band gap of the FPnCs. Under this condition, the wave becomes trapped in the resonator and decays evanescently through the FPnCs, coupling weakly adjacent resonators \textcolor{blue}{\cite{ManjarrezMontanez2025}}. 
Consequently, the dynamics of the elastic waves along the chain is governed by the tight-binding model with the Hamiltonian in Eq.~(\ref{Eq.Hamiltoninan}), where the energy parameters are replaced by frequencies \cite{LopezToledo2022}. 
The hoppings are geometrically controlled through the width $c_i$ that appears in the FPnCs that connect adjacent resonators. 
These widths were systematically varied to realize three distinct cases with ratios $c_1/c_3=0.93, 1$, and $1.15$, systems I, II and III, respectively. The numerical simulations were done using the finite-element method.  
The frequency spectrum and normal-mode wave amplitudes of a CRPnM as a function of the ratio $c_1/c_3$ are shown in Fig.~\ref{fig:placeholder}(b). The cases with $c_1/c_3=0.93$ and $1.15$ exhibit gapped phases hosting localized topological edge states, as shown in Fig.~\ref{fig:placeholder}(a) and (c). In contrast, the $c_1/c_3=1$ geometry corresponds to the critical gap-closing condition associated with two Dirac cones. Both topological phases exhibit wave amplitudes that decay exponentially toward the center of the chain. 

\begin{figure*}[tbp]
\includegraphics[width=1.6\columnwidth]{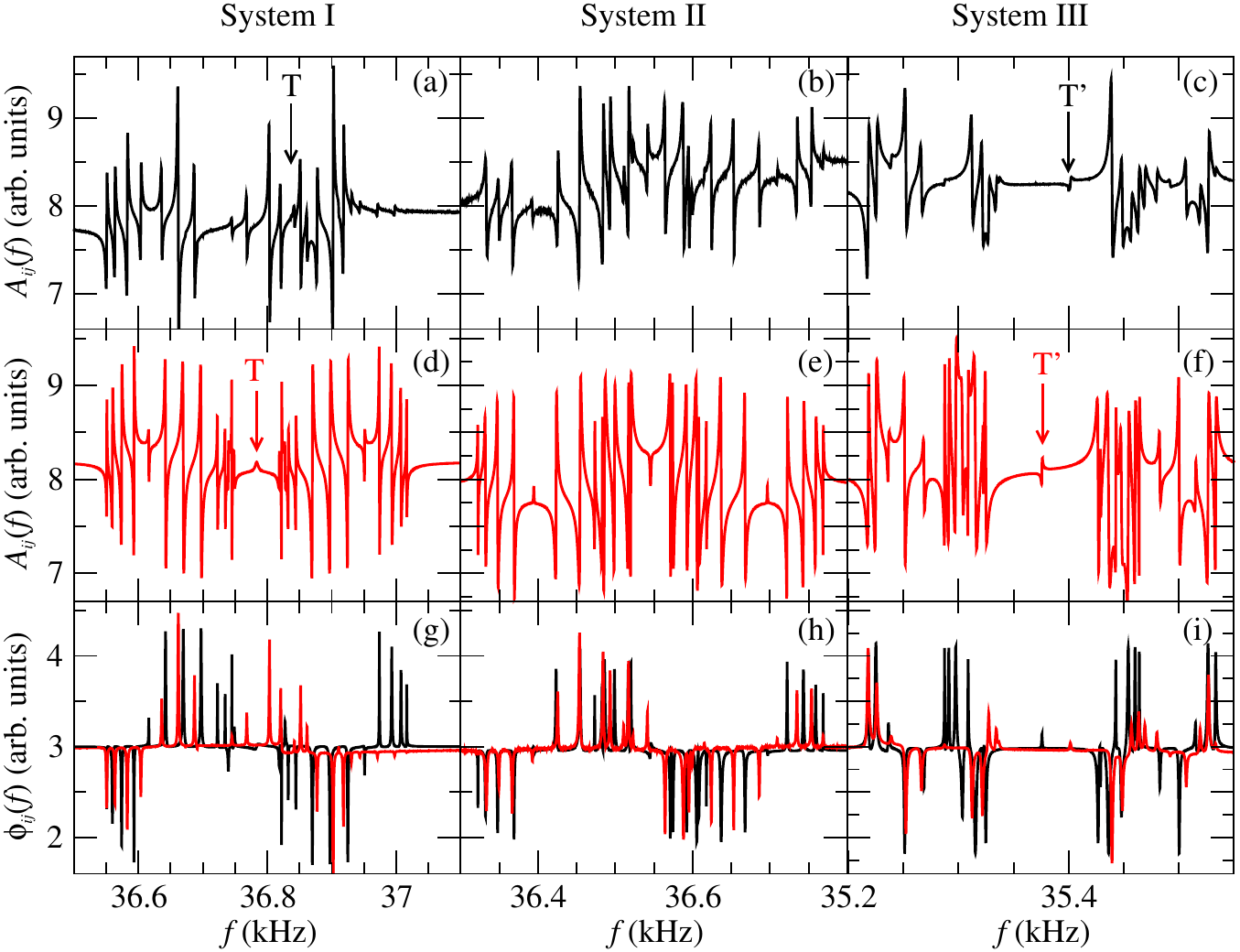}
\caption{Magnitude (upper panels) and phase (lower panels) of measured signals and the corresponding non-equilibrium Green's function prediction (middle and lower panels) for the artificial cis-polyacetylene chain, as a function of the frequency. In panels (a), (d) and (g), $\rho < 1$, in (b), (e) and (h) $\rho = 1$, and in (c), (f) and (i) $\rho > 1$. For $\rho < 1$, a gap opens, showing a first nontrivial phase with a topological state marked by the letter T in panels (a) and (d). This gap closes at $\rho = 1$ see panels (b) and (e). The gap reopens when $\rho > 1$, where a second non-trivial phase appears see panels (c) and (f)); the topological state is marked by the letter T'. 
The panels (g), (h), and (i) give the phase, Eq.~\eqref{eq:Quantities}). The theoretical and experimental results correspond the red and black lines, respectively.}
\label{fig:CisAll}
\end{figure*}

The experimental characterization of the systems was performed using a vector network analyzer {Anritsu} MS4630B combined with contactless electromagnetic acoustic transducers (EMATs) for excitation and detection (See Fig.~\ref{setup}(a)).  
Since the in-plane modes exhibit maximum deformation at the resonator corners, both transducers were positioned along the lateral boundaries of the structure in Fig.~\ref{setup}(a), maximizing the excitation and detection efficiency of the in-plane elastic response. 
Three aluminum systems (Young modulus of 70~GPa, Poisson ratio of 0.33, and density mass of 2700~{kg/m$^3$}) were fabricated as monolithic structures to verify the predicted topological phases. 
During the measurements, the excitation was done in different locations, while the detected response was systematically recorded across the chain. 
Experimental spectra reveal frequency bands for the three fabricated systems with different coupling ratios $c_1/c_3$. 
In the critical regime \(c_1/c_3=1\), the elastic response is concentrated within the $36.30-36.77$~kHz interval, where the band gap closes, and no edge-localized states are observed. 
This corroborates the existence of the Dirac critical points separating the two topological phases, as shown in Fig. \ref{fig:CisAll}(b). 
Away from the critical point, the opening of the gap gives rise to two distinct topological phases supporting localized edge soliton states. 
For the \(c_1/c_3=0.93\) configuration, the bands extend from \(36.53\) to \(37.03\) kHz and exhibit a bandgap of approximately $32$~Hz, within which a localized state emerges near $36.83$~kHz (case (a) in the same figure). 
In the same way, the $c_1/c_3=1.15$ phase exhibits two bands in the interval between \(35.24\) and \(35.56\) kHz with a localized state around $35.43$~kHz, within the gap (see Fig.~\ref{fig:CisAll} (c)). 
In the same figure the phases of the measured signal are given by the black line in panels (g), (h), and (i) for systems I, II, and III, respectively. 
The measured spectra capture the bandgap closure at the Dirac critical point and the emergence of localized edge states upon gap reopening in both topological phases. 

To obtain a theoretical prediction for the magnitude and phase of the measured signal we apply the tight-binding model for the finite artificial cis-polyacetylene, equation~(\ref{eq:HCis}), along with the NEGF method, equations~(\ref{eq:Green}) and (\ref{eq:Quantities}). The parameters $f_{0}$, $t_{1}$, $t_{3}$, $\gamma$, $B$, $C$, $D$, and $F$, were determined from the experimental spectra. 
The central frequency $f_0$ corresponds with the center of the band, while the hoppings, $t_{1}$ and $t_{3}$, are obtained by adjusting those parameters to the experimental bandwidth under the conditions that $\rho < 1$, $\rho = 1$, and $\rho > 1$. 
The coupling strength to the outside, $\gamma$, is obtained by a fitting to the width of the first resonance for each spectrum.  
Finally, the parameters $B$ and $D$ are scaling factors of the measured signals, while $C/F$ are displacements along the $A(f)/\phi(f)$ axis, which are straightforwardly obtained from the spectra. The values of the relevant parameters are summarized in Table~\ref{tab:parameters}. 
The exciter and detector are located on the sites ($i,j$): (1,6), (1,6), and (6,10), for the cases $\rho < 1$, $\rho = 1$, and $\rho > 1$, respectively.

In Figure~\ref{fig:CisAll}, the comparison between the measured signals (black lines) and the theoretical predictions of equation~(\ref{eq:Quantities}) (red lines) is depicted in magnitude and phase. 
Firstly, when $\rho < 1$, a gap opens with the appearance of a first nontrivial phase where the frequency corresponding to the topological state is marked by a black letter T, see panels (a), (d) and (g).  
Secondly, when $\rho = 1$, the gap closes and two non-equivalent cones show up at the center of the band, see panels (b), (e) and (h). 
Finally, when $\rho > 1$ the gap reopens and a second non-trivial phase appears, see panels (c), (f) and (i). 
The topological protected state is indicated by a T/T' letter for the experiment/theory. 
The experimental results present excellent agreement with numerical simulations and with the theoretical predictions of the twofold topological phase transition model with long-range hoppings.  
Although some differences are observed between experiment and theory, the discrepancies in frequency are below 0.2\%. 
They come from different sources as machining errors on the plate, effects of second-nearest neighbors and non-orthogonal orbitals, among others. Also, as it can be seen in Fig.~\ref{fig:CisAll}, the bands change its position for different values of $c_1/c_3$. Despite of that, the edge states remain topologically protected. 

\begin{figure*}[tbp]
    \centering
    \includegraphics[width=1.6\columnwidth]{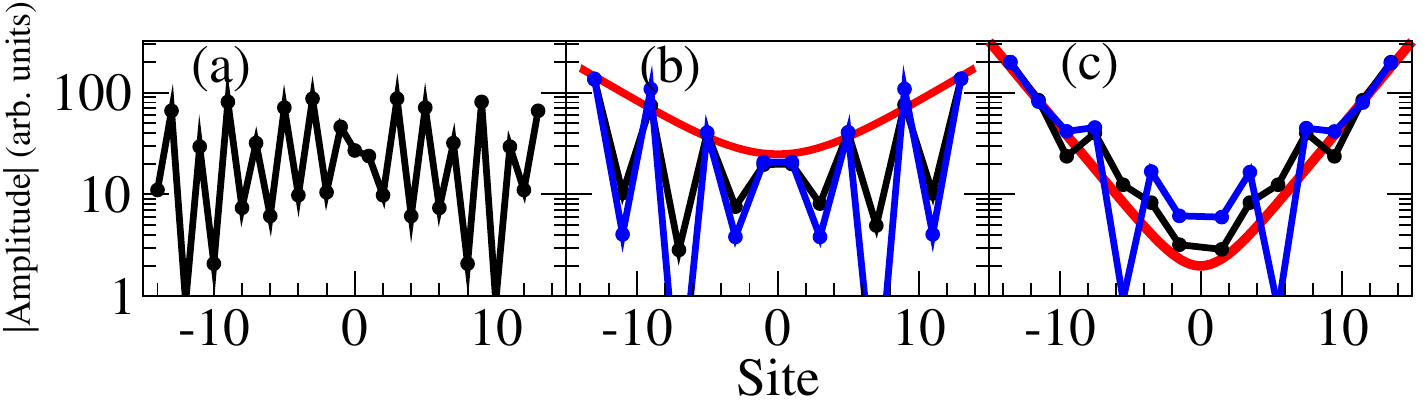}
    \caption{(a) Spatial profile of the critical mode at \(c_1/c_3=1\). The nearly uniform amplitude distribution indicates the delocalized character of the mode at the Dirac gap-closing point. Experimental (black) and numerical (blue) spatial amplitude profiles of the topologically protected elastic modes for (b) $c_1/c_3=0.93$ and (c) $c_1/c_3=1.15$ showing an exponential decay towards the center of the chain. Experimental amplitudes are renormalized for comparison with numerical predictions. The lattice sites were labeled from the center of the chain. The res lines correspond to fittings to the envelope of the amplitude.}
   \label{edgeStates}
\end{figure*}

The wave amplitudes were experimentally mapped along the entire chain to characterize its exponential localization and compared with numerical simulations. 
In the critical case, a mode near the center of the band was selected to observe its behavior. 
As expected, the measured amplitude extends across the chain completely delocalized (see Fig.~\ref{edgeStates})(a). 
In contrast to the extended modes of the critical phase, the edge states observed in both topological phases exhibit localization at the boundaries of the system, {\sl i.e.}, they show an evanescent decay toward the center of the chain. 
This behavior is directly evidenced by the logarithmic amplitude profiles as a function of lattice site, which display an exponential localization. 
The experimental data were fitted with the exponential function {$\sim 0.4\cosh(x/5.3)$} for the case in Fig.~\ref{edgeStates}(b) and {$\sim 0.02\cosh(x/2.6)$} for the case in Fig.~\ref{edgeStates}(c). 
These results confirm the localized nature of the observed wave amplitudes as solitons and are consistent with the predictions of the model in \cite{BetancurOcampo2024} with long-range hoppings.

\begin{table*}[t]
\centering
\vspace{1.5ex}
\begin{tabular}{c c r r c r c r c}
\hline\hline
\noalign{\smallskip}
 \phantom{A} Regime \phantom{A} & $f_{0}$~(kHz) & \multicolumn{1}{c}{$t_{1}$ (Hz)} & \multicolumn{1}{c}{$t_{3}$ (Hz)} & $\gamma$ (Hz) &  \\ [0.5ex]
\hline
\noalign{\smallskip}
$\rho > 1$ & 36.78 & $-92.15$ & $-51.45$ & 7.5  \\ 
$\rho = 1$ & 36.55 & $-75.53$ & $-75.52$ & 7.5  \\
$\rho < 1$ & 35.38 & $-36.78$ & $-86.57$ & 7.5  \\
\noalign{\smallskip}
\hline\hline
\end{tabular}
\caption{\label{tab:parameters} Parameters used in the NEGF method, obtained by fitting the experimental data.}
\end{table*}

To conclude, we have experimentally verified the twofold topological phase transition in a one-dimensional chain with first- and third-nearest-neighbor hoppings using coupled-resonator phononic molecules. 
By systematically detuning the phononic bond geometry, we observed bandgap closures, inverted gaps, and robust edge states. 
Our results bridge theoretical predictions and practical realizations, establishing coupling-resonator phononic molecules as a powerful platform for exploring complex topological phenomena in one dimension.
The ability to carve valley-pseudospin textures and control mass asymmetry opens prospects for one-dimensional valleytronics and directional waveguiding. 
Beyond reproducing the physics of the extended SSH model, the proposed system enables robust control of elastic modes and provides a versatile platform for the experimental exploration of topological wave phenomena in engineered mechanical structures. 
Our use of coupled-resonator phononic molecules demonstrates a versatile route to engineer long-range hopping in phononic systems.
The universal nature of the transition is supported by reproducing electronic-scale predictions at kilohertz frequencies, consistent with Bloch's theorem-based scale invariance.

\section{Acknowledgments}
 B.M.-M. and J.A.F.-M. acknowledge ﬁnancial support from SECIHTI. A.M.M.-A. and E.F.-O. acknowledge financial support from SECIHTI under program ``Estancias Posdoctorales por M\'exico 2022''. This work was supported by DGAPA-UNAM under projects PAPIIT IN108825 and IA102125.

\bibliography{Twofold_exper}
\end{document}